**Preliminary structural study of chromium coatings for nuclear applications**

*Michał A. Stróżyk*, Jong-Dae Hong, JaeYong Kim, Iwona Jóźwik, Ryszard Diduszko, Jerzy Zagórski, Witold Chromiński, Kinga Suchorab, Katarzyna Ciporska, Marcin Brykała, Jacek Jagielski*

*E-mail: Michal.Strozyk@ncbj.gov.pl

M. A. Stróżyk, I. Jóźwik, R. Diduszko, W. Chromiński
NOMATEN Centre of Excellence, National Centre for Nuclear Research, Andrzeja Sołtana 7, 05-400 Otwock-Świerk, Poland

J. D. Hong, J. Y. Kim
LWR Fuel Technology Research Division, Korea Atomic Energy Research Institute, 989-111 Daedeok-daero, Yuseong-gu, Daejeon, 34057, Republic of Korea

J. Zagórski, K. Suchorab, K. Ciporska, M. Brykała, J. Jagielski
Materials Research Laboratory, National Centre for Nuclear Research, Andrzeja Sołtana 7, 05-400 Otwock-Świerk, Poland

W. Chromiński
Warsaw University of Technology, Faculty of Materials Science and Engineering, Wołoska 141, 02-507 Warsaw, Poland

**Funding:**
This project has been supported by NCBiR and KETEP grant ATF Cladding, grant agreement (in Poland) DWM/POLKOR/2/2024.

**Abstract**

Following the Fukushima Daiichi disaster, an increasing number of studies concentrate on the development of Accident Tolerant Fuel (ATF) cladding materials for nuclear fuel, aiming to prevent the oxidation of zirconium during incidents such as Loss of Coolant Accident (LOCA) and to effectively lower the amount of heat and hydrogen released during emergency core cooling (ECC). Zirconium alloy cladding with a protective chromium (Cr) coating is considered one of the promising candidates, largely due to its relatively short timeline for deployment in nuclear power plants.

In this study, Cr-coated and uncoated Zircaloy-4 claddings were evaluated using high temperature X-ray diffraction (HT-XRD) in vacuum over a temperature range from RT to 1100ºC. The temperatures corresponding to the formation of oxide phases are >200ºC and >600ºC for the uncoated and Cr-coated samples, respectively. SEM and TEM characterisation of the sub-surface in Cr-coated specimen after HT-XRD revealed Fe segregation, formation of $Zr(Fe,Cr)_2$ Laves phase and nano-bubbles at the former Cr / Z4 interface.

## 1. Introduction

As a result of the Fukushima Daiichi disaster, global research efforts intensified around the development of Accident Tolerant Fuel (ATF) systems [1]. These new types of nuclear fuel cladding materials are designed to prevent the oxidation of zirconium during severe incidents like a Loss of Coolant Accident (LOCA), reduce cladding degradation, thickness changes, and lower the amount of heat and hydrogen released during emergency core cooling (ECC) procedures [2,3]. Zirconium-based alloys have been used as nuclear fuel cladding materials since 1950s and therefore, surface coated Zr-based claddings are considered the most promising near-term candidates, with relatively short timeline for deployment in nuclear power plants.

Various protective coating options have been proposed so far, including pure Cr [4–7], CrAl [8,9], FeCrAl [10], $Cr_2AlC$ [11,12], $Ti_2AlC$ [13], etc. Among different plating techniques, the Arc Ion Plating (AIP) has been identified as a reliable method for deposition of Cr [14] or Cr-alloy/FeCrAl coatings [15], showing good adhesion on Zr-1.1Nb or Zircaloy-4 cladding materials. Moreover, Cr-based coatings could potentially reduce the amount of crud deposited during nuclear power plant operation, making them good candidates for cladding applications in more aggressive environments of future fission reactors [9].

In this work, a preliminary structural study of Cr-coated Zircaloy-4 (Z4) cladding material is presented. Oxidation behaviour of both Cr-coated and uncoated Z4 specimens is evaluated using high temperature X-ray diffraction (HT-XRD) over a temperature range from RT to 1100ºC. The temperature ranges corresponding to the formation and transformation of oxide phases are determined. After HT-XRD, the development of surface layers is carefully examined using SEM and TEM. The study aims to develop a comprehensive method of material characterisation for potential ATF applications.

## 2. Experimental Methods

*Sample preparation by Arc Ion Plating (AIP) process:*

A 10 ±2 μm Cr coating was deposited on the commercial CWSR (cold-worked and stress-relieved) Zircaloy-4 (Z4) cladding using AIP method under controlled vacuum conditions. For sample preparation, the Z4 cladding was cleaned ultrasonically in acetone and ethanol. The cleaned cladding was then dried thoroughly and loaded into a vacuum chamber. Before deposition, the Z4 cladding tubes were placed in a vacuum chamber on rotating holders in each system, where they were cleaned by ion etching in argon (Ar) plasma. This process removed thin Zr-oxide and other impurities, thereby improving the adhesion of the Cr coating. The AIP coating process was then performed to grow a Cr layer using a pure Cr target (99.9%) below 200°C to prevent microstructural changes of the Zr substrate [8,9].

*Thermogravimetric (TG) Analysis:*

TG analysis was carried out using a Netzsch STA 449 F3 Jupiter equipped with a silicon carbide (SiC) furnace. For the measurements, ring-shaped samples with an outer diameter of 9 mm, wall thickness of 1 mm and height of 5 mm, were positioned on an alumina plate placed on a TG pin carrier. The TG analysis was carried out in two atmospheres - (1) air with a flow rate of 125 mL $min^{-1}$ and (2) water vapour with a flow rate of 125 mL $min^{-1}$ and relative humidity of 80 – 90%. Calibration of the device was performed over a broad temperature range using reference materials (In, Sn, Bi, Zn, Al, Au). All measurements were carried out in the temperature range of 20–1100°C, at a heating rate of 5°C $min^{-1}$.

*Optical micrographs:* Optical micrographs were acquired using a Keyence VHX 7000 digital microscope equipped with a VHX-7020 camera with a 2K lens (20-200x). Observation od samples after thermal analysis was performed at 20x magnification.

*SEM characterisation*: SEM imaging, EDS and EBSD measurements were performed using a Helios 5 UX (ThermoFisher Scientific) dual beam microscope, equipped with Elstar™ field emission electron column, Ga+ Phoenix™ Focused Ion Beam (FIB), EDAX Elite Super (EDS) and EDAX Velocity Pro (EBSD) systems. Cross-sections and lamellas for TEM were prepared using the FIB lift-out technique operated at 30 kV. A Pt protective layer was deposited on top of the selected area as part of the lift-out procedure. The final thinning of the sample was performed at 5 kV and 2 kV to minimize the damage layer thickness introduced by the $Ga^+$ ions.

*TEM characterisation*: TEM observations were performed using a JEOL JEM F200 microscope (accelerating voltage of 200 kV) operated in a STEM mode for imaging and EDS analysis (two Centurio XXL detectors with total collection angle of 1.7 sr), and in a TEM mode for selected area electron diffraction (SAED).

*High temperature X-Ray diffraction (HT XRD):* The samples were placed in a high-temperature chamber on a corundum table and a corundum support in the case of the Cr/Z4 sample, or on a "backgroundless" Si510 single crystal plate in the case of the reference Z4 zirconium alloy sample. Coupled θ-2θ measurements in Pb-GM geometry were performed using a Bruker D8

Advance diffractometer. The Cu anode was powered at 40 kV and 40 mA ($\lambda_{CuK\alpha1}$ ≈1.5406 Å). 0.6 mm primary fixed divergent slit (FDS), 10 mm vertical mask, 0.3° secondary equatorial Soller and 2.5° primary and secondary axial Soller slits were used. The LYNXEYE XE-T detector working in the high energy resolution and 0D modes without Ni-filter was used. In-situ high temperature HT-XRD measurements were carried out using an Anton Paar HTK 1200N chamber in the temperature range from 25°C to 1100°C, namely at 25°C, 200°C and then every 100°C until 1100°C, additionally at 850°C. All scan covered 2θ range of 20 – 148° with 0.01° step. Specular reflection coupled θ-2θ scans were measured, time per step 0.28s, two diffraction patterns acquired at each temperature step. The total time of diffraction patterns measurements at one constant temperature was 2 h. The vacuum in the range of $10^{-5}$-$10^{-6}$ mbar was maintained during HT-XRD. The Bruker DIFFRAC.EVA program with the database of diffraction standards ICDD PDF5+ 2025 [16] and DIFFRAC.TOPAS programs were used for phase analysis and to refine the models of the identified phases to describe the experimental diffraction patterns [17–19]. The instrumental peak broadening was assessed using the $LaB_6$ NIST SRM 660b and the same diffractometer geometry.

**3. Results and discussion**

3.1 Deposition of Cr layers

As shown in **Figure 1B**, the thickness of the as-deposited Cr coating using AIP method is 10 ±2 µm. The columnar grains of Cr coating are aligned along the radial direction of the cladding. Meanwhile, due to the nature of the AIP process, significant local heat generation can occur, making it crucial to ensure that the Z4 substrate does not undergo changes that could compromise the characteristics of the cladding. To verify this, microstructural observations were conducted using visible light microscopy, and it was confirmed that the stress relief annealed (SRA) microstructure of Z4 substrate material was maintained after the coating process (**Fig. 1A vs 1C**).

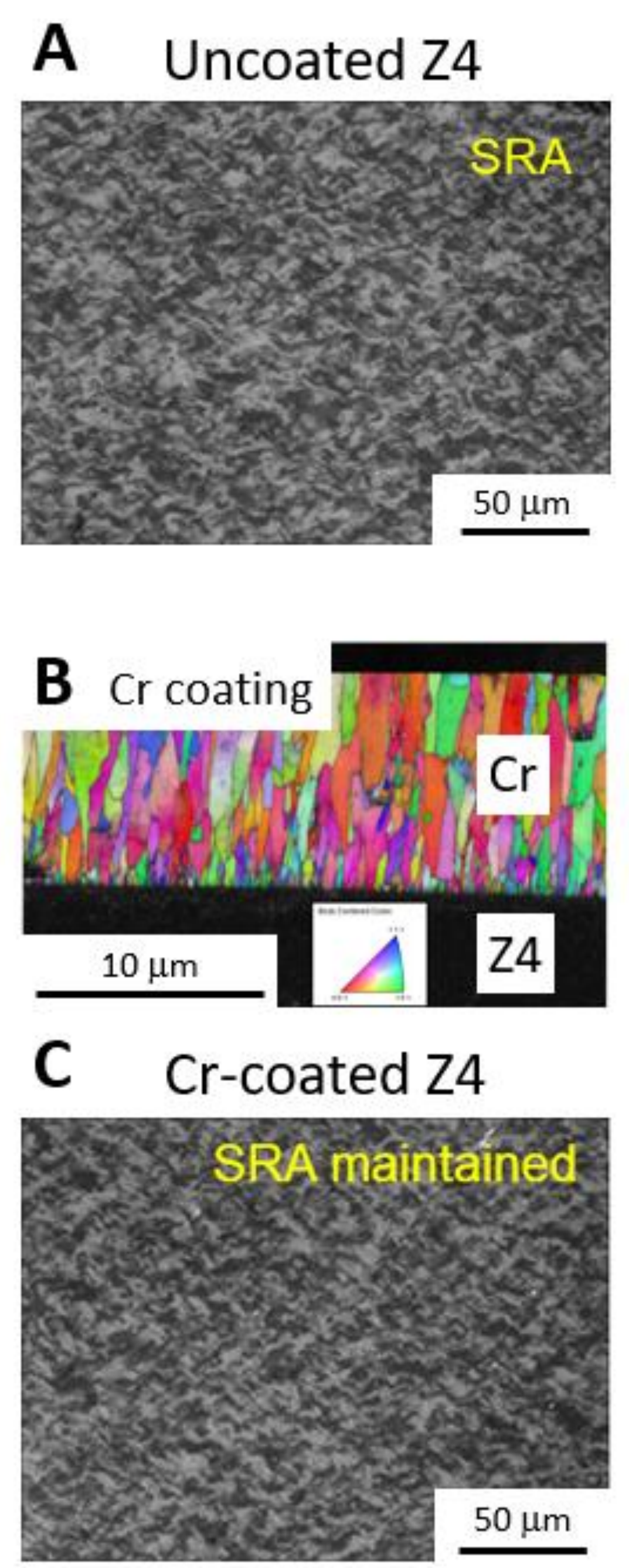

**Figure 1.** Cross-sectional images of the Cr coating and Z4 alloy substrate: A) Stress relief annealed (SRA) microstructure of Z4 alloy before the Arc Ion Plating (AIP) process (visible light microscopy); B) Microstructure of the Cr coating by AIP (EBSD – IPF Z map); C) SRA microstructure of Z4 alloy after AIP process (visible light microscopy).

3.2. Structural analysis of Cr-coated samples

3.2.1 SEM analysis prior to HT-XRD

**Figure 2** presents visible light and SEM micrographs of the Zircaloy-4 (Z4) cladding rod samples in uncoated and Cr-coated states. It is observed that the Cr layer deposition, results in the visible gloss change (**Fig. 2A vs 2B).** SEM images show no signs of cracking or delamination of Cr layer, neither caused by handling or machining of the rods into smaller samples. SEM images of the uncoated Z4 sample reveal grooves on the surface (**Fig. 2C**), parallel to the circumference of the fuel cladding rod, most likely resulting from surface treatments such as ion etching. Cr coating covers the surface uniformly (**Fig. 2D**), forming slightly corrugated surface and partially eliminating the grooves. Randomly distributed circular nubbles, with diameters ranging from about 10 μm down to the submicron scale, are found embedded in the new surface. Most likely, the nubbles are deposited droplets of molten material (Cr), which are a typical feature for the AIP coating process [15].

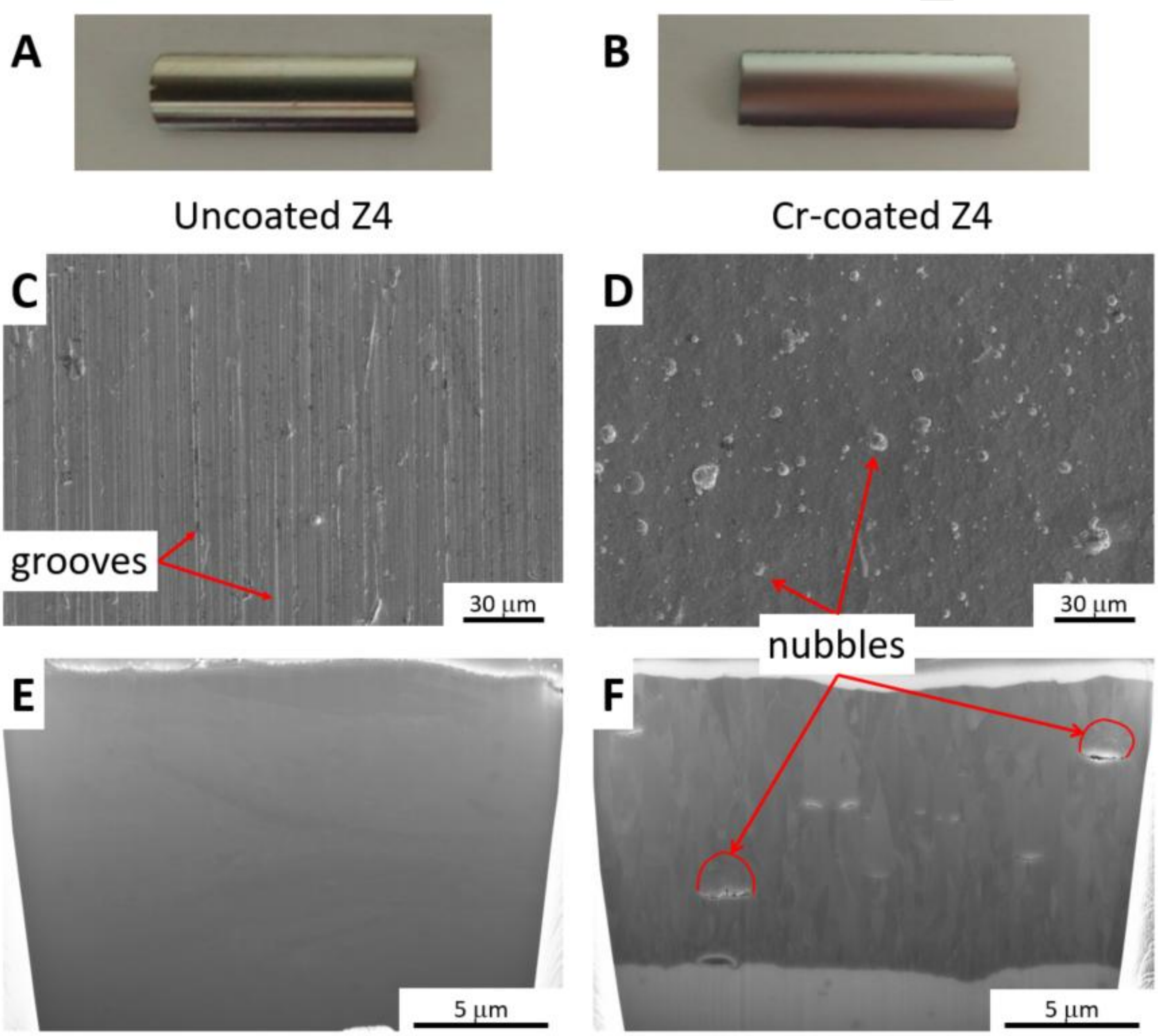

**Figure 2.** Z4 cladding sample with and without Cr coating: visible light microscope (A, B) and SEM images of the surface (C, D) and cross-sections (E, F).

Cross-sections through the surface of the Cr-coated and uncoated Z4 samples, prepared via FIB, are shown in **Figures 2E and 2F**. In general, the protective Cr coating thickness is ranging from 10 to 11 μm. The grains of the Cr layer are elongated in the growing direction (radial direction of the cladding) and characterised by a coarser grain size compared to underlying Z4 cladding material. The presence of nubbles and pores is also observed within the Cr layer (**Fig. 2F)**. The occurrence of pores is directly related to the presence of nubbles, because in most cases they are located directly under the nubbles (occurring both on and below the surface), with a characteristic grain refinement above the crack being a trace of the molten Cr presence (cross-section image in **Fig. 2F**). No other cracks or delamination of the Cr layer are observed in cross-sections of the tested samples.

3.2.2 Thermogravimetric Analysis

TG measurements were conducted at 1100 °C on uncoated Zircaloy-4 (Z4) claddings and chromium-coated claddings (Cr/Z4), as shown in **Figure 3**. The TG curves present the mass gain as a function of time: in a steam atmosphere (left), and in air (right). These experiments revealed significant differences in the behaviour of uncoated (Z4) versus chromium-coated claddings (Cr/Z4), as well as the effect of the surrounding atmosphere. The oxidation kinetics of both examined materials proceeds in two stages: an initial sub-parabolic mass increase corresponding to the formation of a dense and protective $ZrO_2$ layer on the surface, followed by a rapid acceleration of oxidation after the breakaway caused by cracking and/or spallation of the oxide, allowing the direct access of the oxidising media to the metal. The recorded mass gains for Z4 is approximately 12 % in steam and 17.6 % in air, whereas Cr/Z4 exhibits lower mass gains of roughly 6.6 % and 7.7 %, respectively. On the TG curves of samples oxidized in air, a higher mass gain is observed compared to the water steam environment. Similar results were previously reported [20]. This difference in behaviour between the samples comes from the presence of nitrogen in air, that accelerates corrosion [21]. Macroscopic differences in surface degradation are clearly visible in the visible light microscope images shown in the upper part of the graphs in **Fig. 3**. Uncoated claddings exposed to steam exhibit an orange oxide layer on the edges, that is non-adhesive and flaky – a typical effect of oxide volume expansion, leading to stress accumulation and fragment detachment. In air, the orange, non-protective $ZrO_2$ layer is also present, but appears more compacted and less fragmented. This difference in behaviour between the samples comes from the lower hydrogen content in the air atmosphere. At higher hydrogen concentration in the steam, hydrogen could diffuse into the material (hydrogen pickup), resulting in hydrogen-inducted embrittlement [22]. In contrast, Z4 claddings coated with chromium (Cr/Z4, red lines in **Fig. 3A and 3B**) display significantly lower mass gains up to 1000°C. Photographs reveal a black, continuous, compacted, and intact surface in both steam and air. After measurements conducted in steam, the surface of the coating becomes matte and dark grey, whereas in air it preserves a slight metallic sheen, which may indicate the formation of a thin, protective $Cr_2O_3$ layer [23]. The matte appearance may result from a highly developed surface and the presence of numerous defects like micropores, microcracks, and worm-like structures with whisker-like growths [23]. Additionally, after oxidation in both atmospheres, curled edges are observed in Cr-coated samples. This effect is attributed to the oxidation and subsequent expansion of the uncoated zirconium surface exposed during cutting. Presented results prove, that the Cr layer acts as a barrier, limiting substrate oxidation and suppressing zirconium oxide growth and cracking. As a consequence, the chromium coating markedly enhances the corrosion resistance of Z4 at 1100°C by substantially reducing the oxidation rate of the underlying metal. Similar observations were reported in the studies by H. Yeom, where the Cr coating provided about 50 times improvement in oxidations resistance compared to Z4 alloy in steam environment at 1310°C [24].

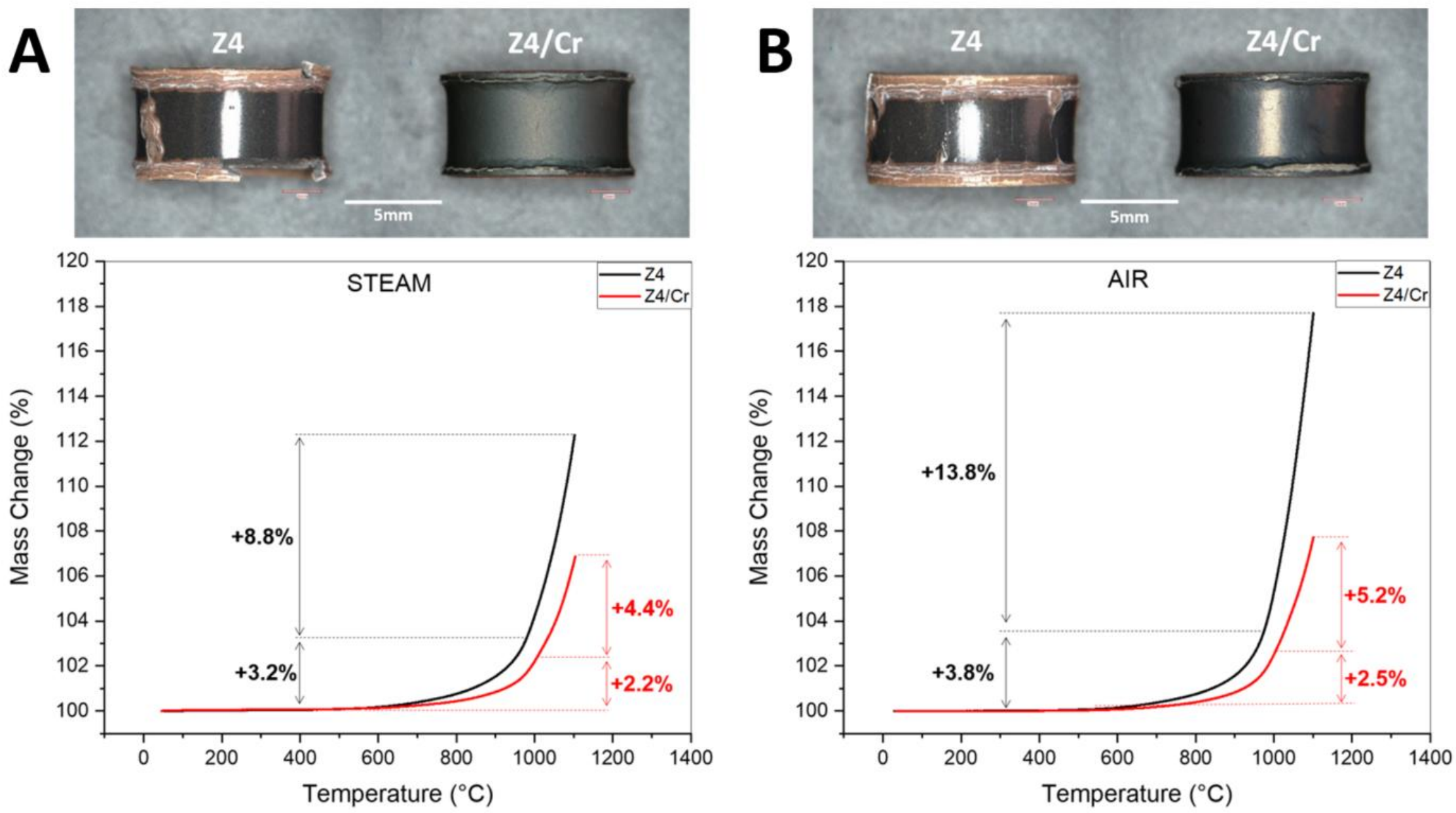

**Figure 3.** Photographs of Zircaloy-4 (Z4) and Zircaloy-4 coated with Chromium (Cr/Z4) cladding segments after thermogravimetric (TG) tests, together with corresponding mass gain curves, during oxidation at 1100°C in steam (A) and in air (B).

3.2.3 High Temperature X-Ray Diffraction

In order to study the high temperature stability of the Cr-coated and uncoated Z4 samples, high-temperature XRD (HT-XRD) was performed. Samples were analysed at room temperature and in a range of 200 – 1100°C at intervals of 100°C. Additional measurement at 850°C has also been performed, due to expected α > β Zr transformation in 800 – 900°C region. **Figure 4** summarises HT-XRD results for the uncoated Z4 (A) and Cr-coated Z4 samples (B). Room temperature measurements confirm hexagonal (HCP) α-Zr phase (lattice constants: a = 3.24 Å;

c = 5.15 Å) in the uncoated sample and BCC α-Cr phase (lattice constant: a = 2.89 Å) in the Cr-coated specimen. For the uncoated sample (**Fig. 4A**), peaks from the monoclinic and tetragonal $ZrO_2$ phases appear between 200 - 300°C. Most likely, the heating element, insulation or surrounding steel of the test chamber may be the source of oxygen, despite high vacuum in range of $10^{-5}$-$10^{-6}$ mbar maintained in the heating chamber during HT-XRD experiments. Reflections of the oxygen-stabilized α-Zr(O) appear between 400 – 500°C, which evidence the presence of a hexagonal solid-solution phase with oxygen ions distributed in structural gaps in the Zr ion lattice. Formation of FCC ZrO high-temperature phase (with a lattice constant of about 4.7 Å), is recorded starting from ~ 850 – 900°C and after cooling of the sample. This phase is probably metastable in RT.

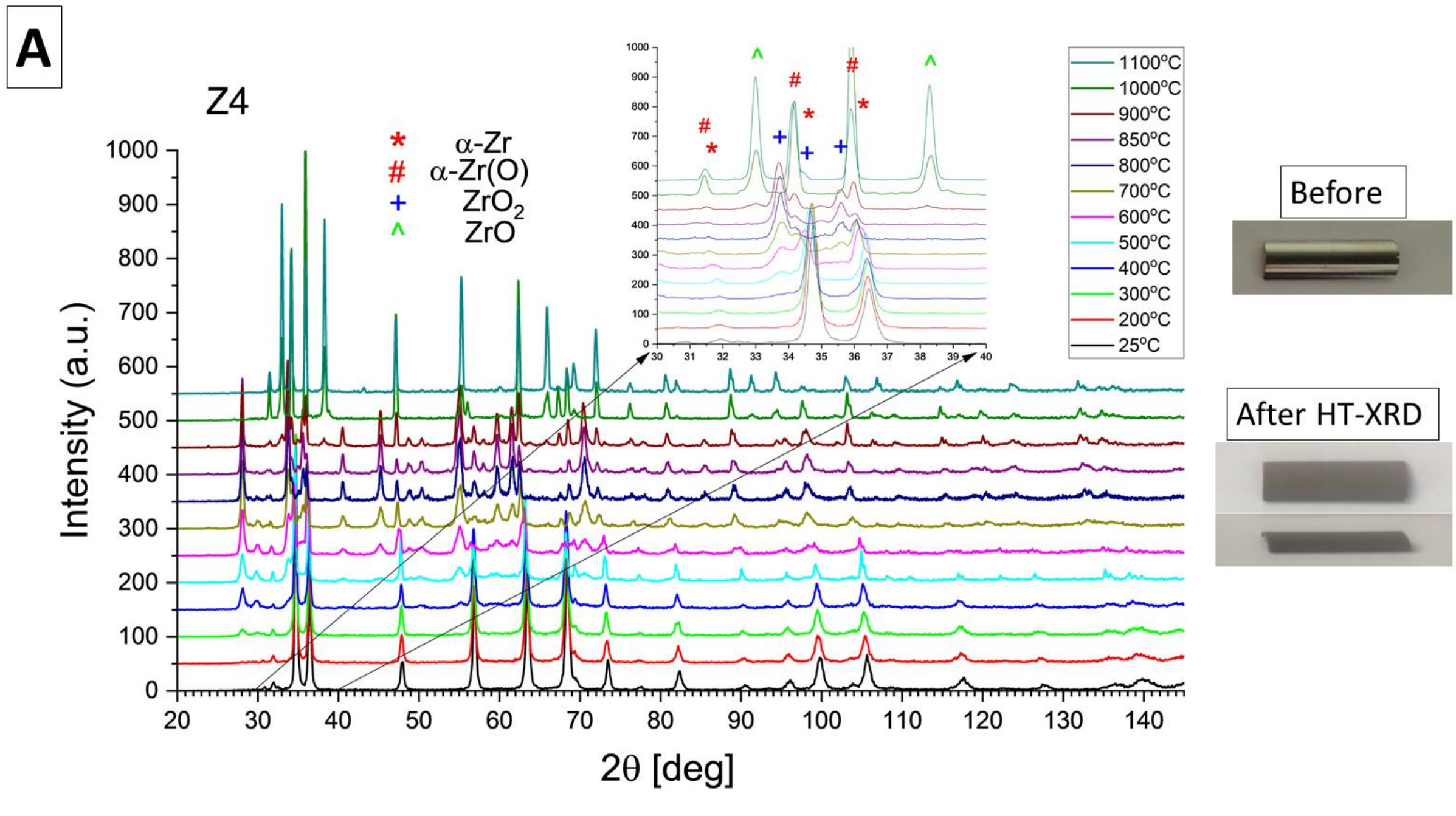

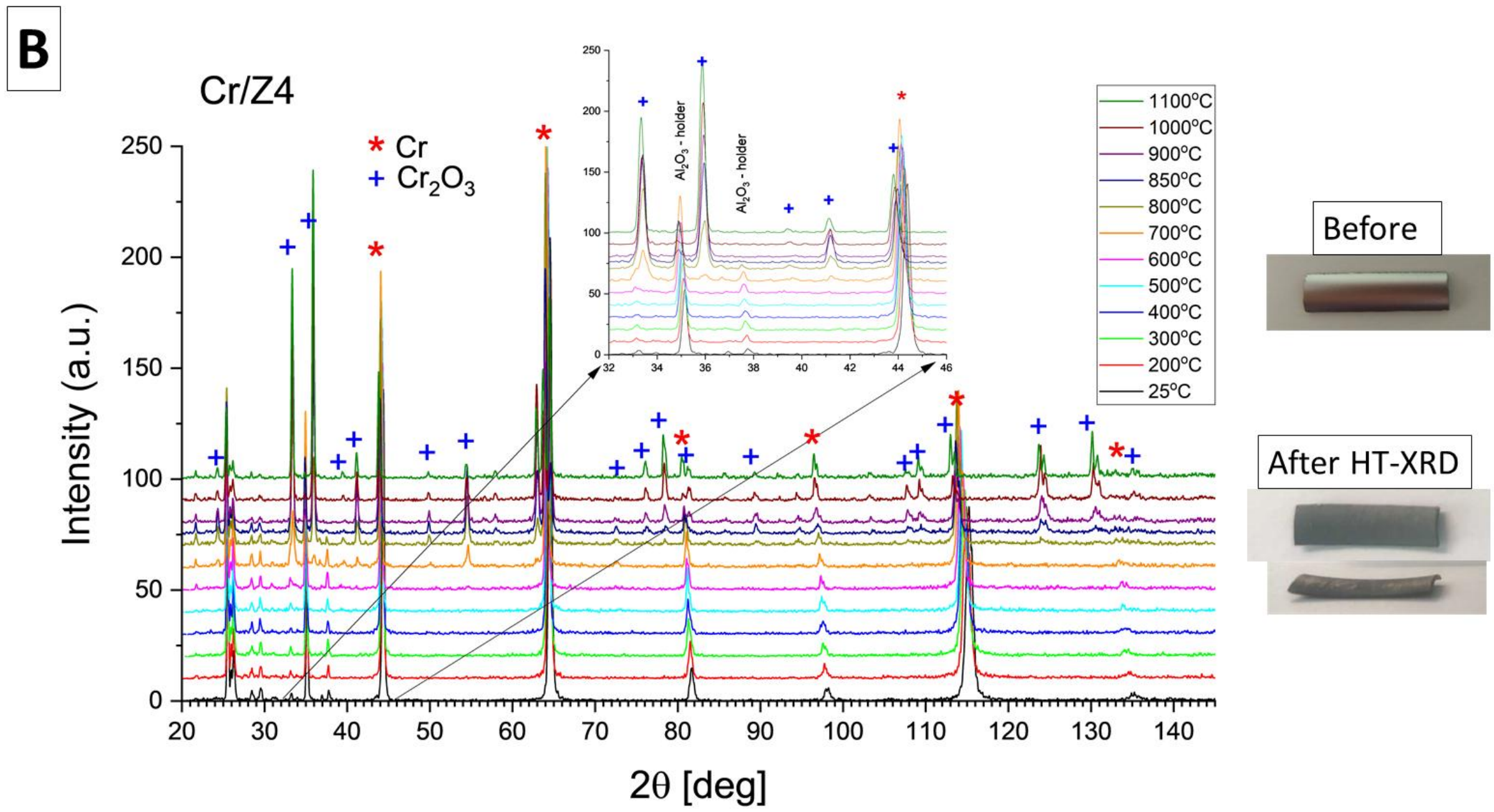

**Figure 4.** HT-XRD of the uncoated and Cr-coated Z4 sample: A) Summary figure of X-ray diffraction patterns for the Z4 sample in the temperature range of 25°C - 1100°C and photographs of the sample before and after HT-XRD; B) Summary figure of X-ray diffraction patterns for the Cr/Z4 sample in the temperature range of 25°C - 1100°C and photographs of the sample before and after HT-XRD.

HT-XRD of the Cr-coated Z4 sample (**Fig. 4B**) revealed that the oxidation of Cr layer begins above 600°C (formation of $Cr_2O_3$ oxide phase), which is significantly higher compared to oxidisation of Zr in uncoated Z4 sample. The $Cr_2O_3$ is thermodynamically more stable than Cr in an oxidising environment and provides a region of lower diffusion kinetics [25]. Below 600°C, only the metallic phase of Cr (BCC phase) is visible in the diffraction patterns. Due to the thickness of the Cr layer (11 μm) and its absorption properties of X-rays, only diffraction reflections from Cr and Cr oxide phases appear using radiation from a Cu tube. Additional, unlabelled reflections visible in the diffraction patterns in the range up to 2θ = 40° come from the corundum plate on which the sample was placed. Moreover, bending of the sample is observed at temperatures above 600°C, which is shown in **Fig. 4B**, most likely resulting from the thermal expansion coefficient differences between the Z4 and forming oxides.

Analysis of diffraction patterns presented in **Figure 4** allowed determination of the lattice constants variations for Cr, $Cr_2O_3$ and Z4, which are plotted in **Figure 5** as a function of temperature. The average linear expansion coefficient determined for the Cr BCC phase (in the direction normal to the sample surface) from **Fig. 5A** is 10.8 $x10^{-6}$ $K^{-1}$. This value is somewhat higher than the values given in international tables [26]: 4.81 $x10^{-6}$ $K^{-1}$ (although tables can be found giving higher values - up to 8 $x10^{-6}$ $K^{-1}$). The above discrepancy is probably related to the occurrence of stresses - the metallic Cr layer is located on the metallic zirconium Z4 alloy and above the temperature of 600°C it is covered with a layer of Cr oxide, with a significant difference in the linear expansion coefficients of these phases. The average linear expansion coefficient determined from **Fig. 5B** for the $Cr_2O_3$ hexagonal phase (in the direction of the radius of curvature of the sample surface) is $\alpha_a$ = 5.7 $x10^{-6}$ $K^{-1}$ for [100] direction and $\alpha_c$ = 7.5 $x10^{-6}$ $K^{-1}$ for [001] direction. Reliable data on the anisotropic linear expansion coefficients for eskolaite ($Cr_2O_3$) was not found, but the volume coefficient is $\alpha_V$ = 18.6 $x10^{-6}$ $K^{-1}$. This structure is similar to corundum, whose volume expansion coefficient is $\alpha_V$ = 23 $x10^{-6}$ $K^{-1}$, and the linear coefficient $\alpha_a$ = 7.3 $x10^{-6}$ $K^{-1}$, $\alpha_c$ = 8.3 $x10^{-6}$ $K^{-1}$ [27], therefore the linear coefficients for eskolaite must be somewhat smaller, close to the experimentally determined values. Finally, the average linear anisotropic expansion coefficients determined from **Fig. 5C** for the Zr hexagonal phase (in the direction of the radius of curvature of the sample surface) is $\alpha_a$ = 9.1 $x10^{-6}$ $K^{-1}$ for 100] direction and $\alpha_c$ = 3.6 $x10^{-6}$ $K^{-1}$ for [001] direction. Most likely, this mismatch of linear expansion coefficients, leads to bending of the Cr/Z4, which was observed above 600°C during HT-XRD experiment (**Fig. 4B**). The above aspects must be taken into account to ensure structural integrity of the future ATF cladding material.

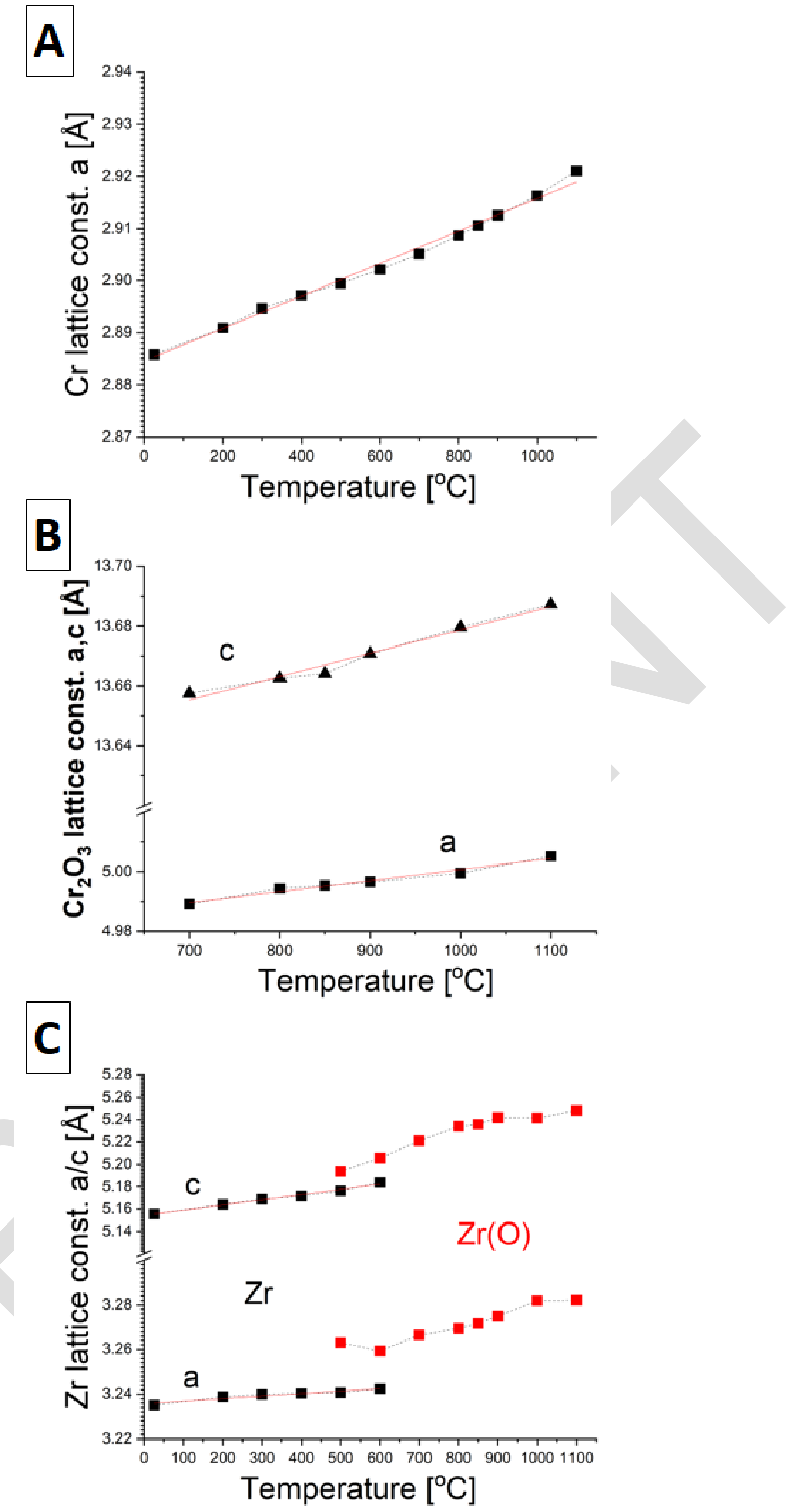

**Figure 5.** Values of the determined lattice constants for: A) Cr lattice constant for the Cr/Z4 sample (in the direction of the radius of curvature of the sample surface) as a function of temperature; B) hexagonal phase $Cr_2O_3$ lattice constants for the Cr/Z4 sample (in the direction of the radius of curvature of the sample surface) as a function of temperature; C) hexagonal phases Zr and solid-solution Zr(O) lattice constants for the Z4 sample as a function of temperature.

3.2.4 SEM and TEM analysis after HT-XRD

After HT-XRD measurements, cross-sections and TEM lamellas were prepared using SEM/FIB technique. The oxide layer, formed during the HT-XRD experiment of the uncoated Z4 sample, is shown on SEM and TEM images in **Figure 6.** The surface layer is identified as ZrO (FCC) high-temperature phase (metastable in RT) through selected area electron diffraction (SAED) in **Fig. 6D**, which is in agreement with XRD measurement performed after cooling of the sample. The pores are readily visible on the surface of the sample (**Fig. 6C**) and are present both in ZrO and α-Zr(O) phases (**Fig. 6A,B**).

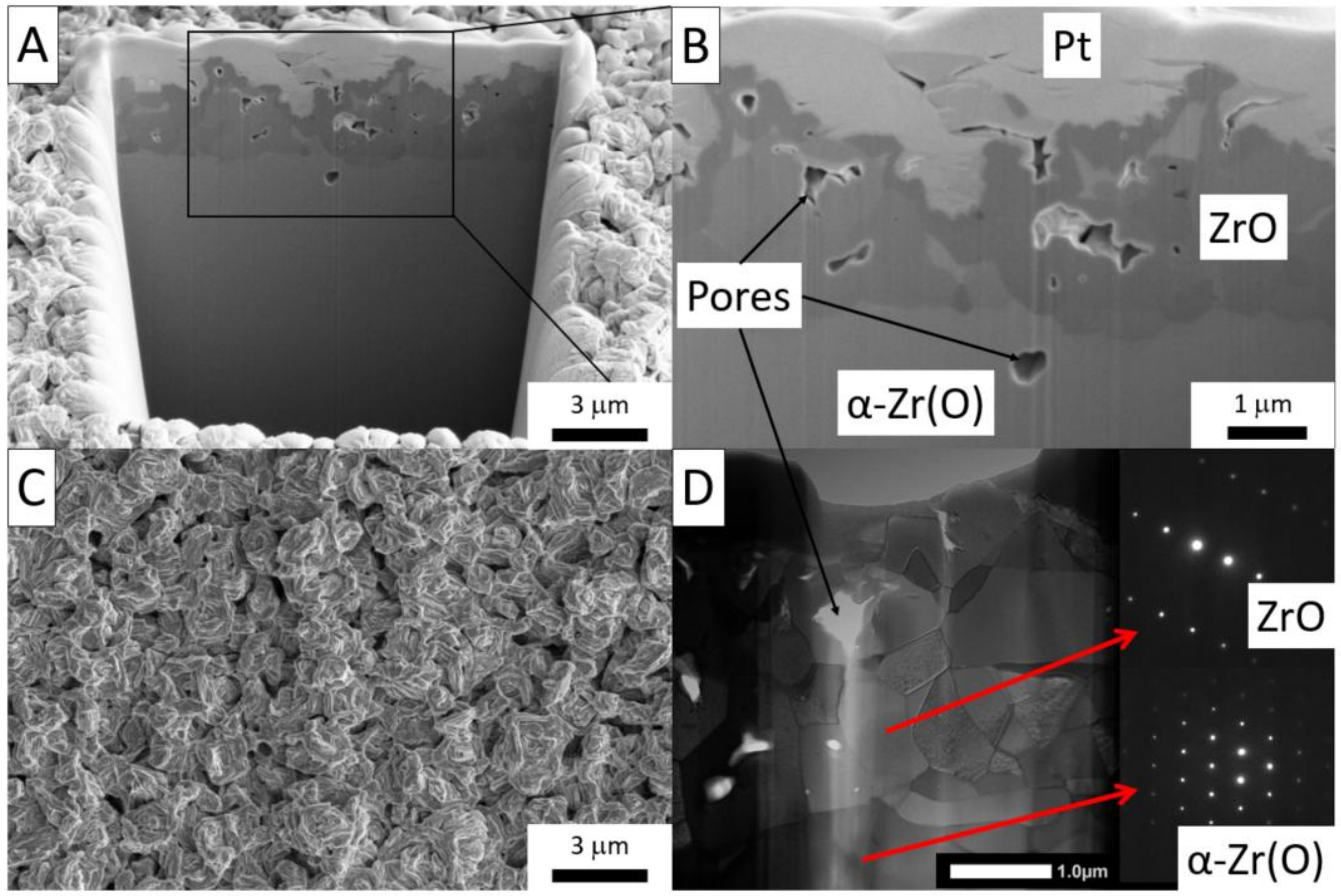

**Figure 6.** SEM and TEM results of the uncoated sample after HT XRD: A) Cross-section of the layers formed during HT XRD; B) SE image showing detail of the cross-section; C) SE image of the top surface after HT XRD; D) Bright-field (BF) STEM image of layers formed during HT XRD, including SAED of ZrO and α-Zr(O) phases.

The surface porosity resulting from HT-XRD is visibly reduced in the Cr-coated Z4 sample shown in **Figure 7**. The oxide forms nearly continuous, but irregular top layer on the sample (**Fig. 7A**). The discontinuities in the oxide expose the underlying BCC Cr phase. Consequently, cross-sections of the sub-surface layer allow observation of at least three distinct zones on top of α-Zr(O) phase material (**Fig. 7B,C**). At the very top (surface) - the $Cr_2O_3$ oxide (eskolaite) is formed, which results in a green tint of the sample after HT-XRD (**Fig. 4B**). The structure of eskolaite is also confirmed by SAED (**Fig. 7D**). As said, this layer covers the sample almost entirely, but regions of discontinuity are readily observed through SEM, as demonstrated in **Fig. 7B,C**, where Cr BCC phase is the top-most layer and the $Cr_2O_3$ disappears. Original chromium coating is thus the second and the thickest layer. Within the layer, severe precipitation of $ZrO_2$ particles is detected, especially at grain boundaries (**Fig. 7C,E**). What is more, a different type of precipitates is present in Cr and $Zr(Fe,Cr)_2$ layers. The careful contrast analysis in SEM images leads to the conclusion that these may be α-Zr(O) [6] (marked with red arrows in **Fig. 7B,C**). Surprisingly, formation of an inter-layer was observed between Cr coating and zirconium alloy (Z4), which may be an intermetallic compound (Laves phase) $Zr(Fe,Cr)_2$, as inferred by SAED (**Fig. 7E**). Similarly as in case of uncoated sample (**Fig. 6**), porosity of the coating can be observed after HT-XRD. The pores can be especially found in the volume of the Cr layer, closer its interfaces with $Cr_2O_3$ (**Fig. 7D**) and $Zr(Fe,Cr)_2$ (**Fig. 7B**).

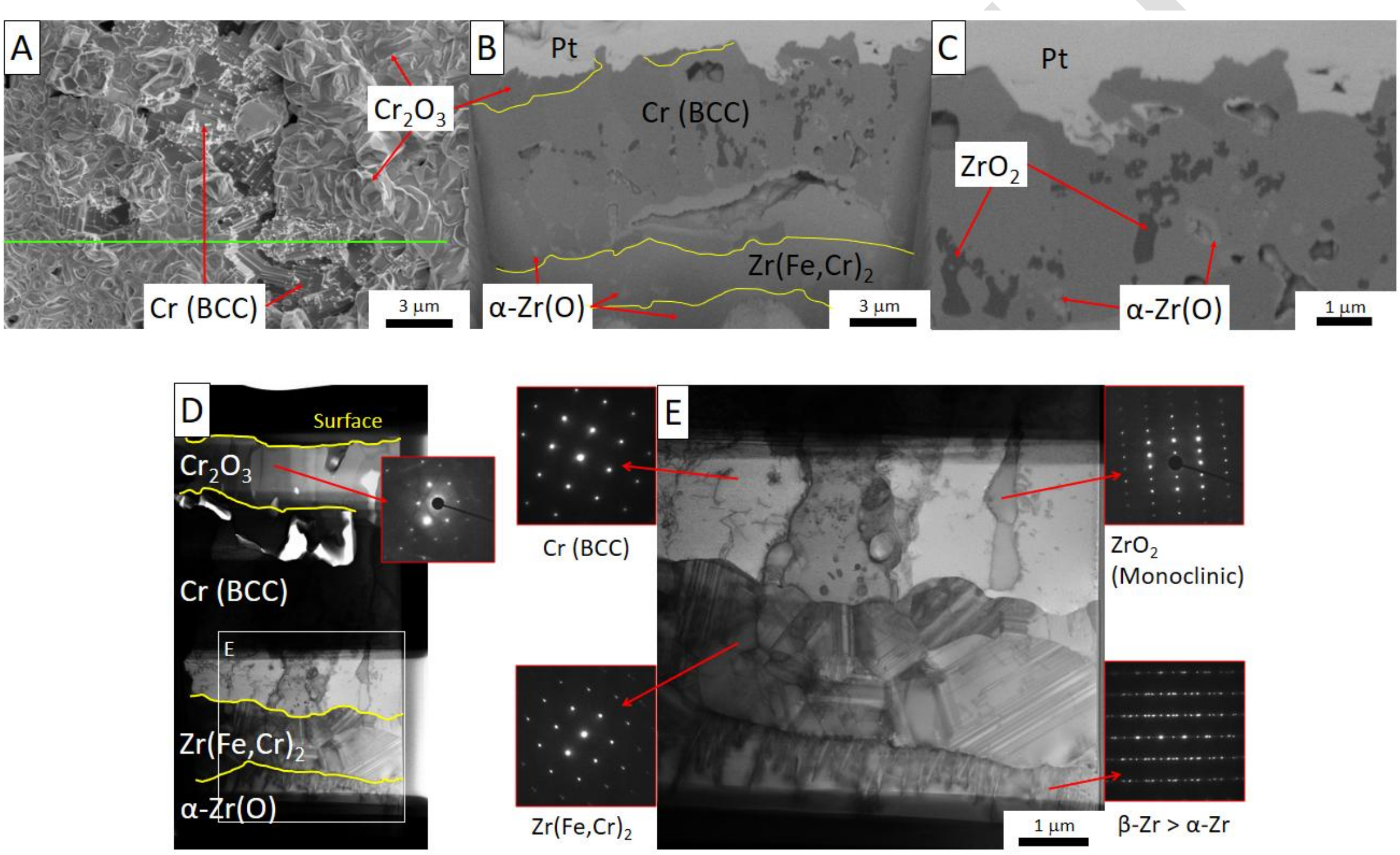

**Figure 7.** SEM and TEM results of the Cr-coated sample after HT-XRD. A) Secondary electron (SE) image of the top surface of the sample showing exact location of the section plane in B (green line) and BCC Cr phase exposed in the discontinuities of the oxide; B) BSE (backscattered electron) image of the cross-section; C) Higher magnification image showing $ZrO_2$ and α-Zr(O) precipitates; D) STEM image of the lamella prepared from the Cr-coated sample after HT-XRD, with SAED of the top $Cr_2O_3$ layer; E) BF STEM (bright field STEM) image and SAED of the layers and precipitates observed at Cr/Z4 interface.

**Figure 8** shows the close-up STEM images and EDS maps of the $Cr/Zr(Fe,Cr)_2/Zr(O)$ interfaces formed during HT-XRD of the Cr-coated Z4 sample. EDS analysis of the interface region (**Fig. 8C**) confirms higher Fe content in $Zr(Fe,Cr)_2$ layer, compared to adjacent layers. Formation of this compound at the Cr/Z4 interface, as inferred by both SAED and EDS analyses (**Fig. 7 and 8**), may be promoted during HT-XRD by α-Zr to β-Zr transformation at elevated temperatures, as previously suggested by Li et al. [4]. It has been noted that the source of the Fe enrichment can be the second phase precipitates (SPP) formed during annealing of Z4 alloy or the Fe present in solid solution (~100 ppm) [5]. Furthermore, it is believed that the Fe segregation at Cr/Z4 interface is favourable in order to lower the interfacial energy [28]. Hence, it tends to be more severe if the misorientation of chromium grains compared to zirconium grains increases.

Nano-bubbles with size in the range of 10 – 30 nm can be observed in $Zr(Fe,Cr)_2$, near or at the interface with α-Zr(O), as shown in **Figure 8B**. Similar cavities were previously reported and may be attributed to Kirkendall effects [6,29,30], where the dissymmetric atomic diffusion of Cr and Zr atoms leads to diffusion of vacancies, coalescence and growth of the cavities. However, they may also be filled with $O_2$. It has been previously postulated that inward oxygen diffusion proceeds through the Cr layer along grain boundaries and formation of $ZrO_2$ precipitates facilitates this process [3,6]. Furthermore, it is generally known and evidenced by HT-XRD of Z4 sample (**Fig. 4A** and **5C**), that above certain temperature, a solid solution phase α-Zr(O) forms, where oxygen ions are distributed in structural gaps in the Zr ion lattice, with maximum oxygen level of about 28 at.% [31]. On the contrary, the solubility of O in β-Zr is only about 10 at.% [32]. Therefore, it is conceivable that the $O_2$, exceeding the solubility limit in β-Zr, forms the nano-bubbles as the result of α-Zr(O) to β-Zr(O) transformation at elevated temperatures. Notably, the level of oxygen also increases the temperature of α-to-β transformation [33], hence the α-Zr(O) is considered to be oxygen-stabilised phase.

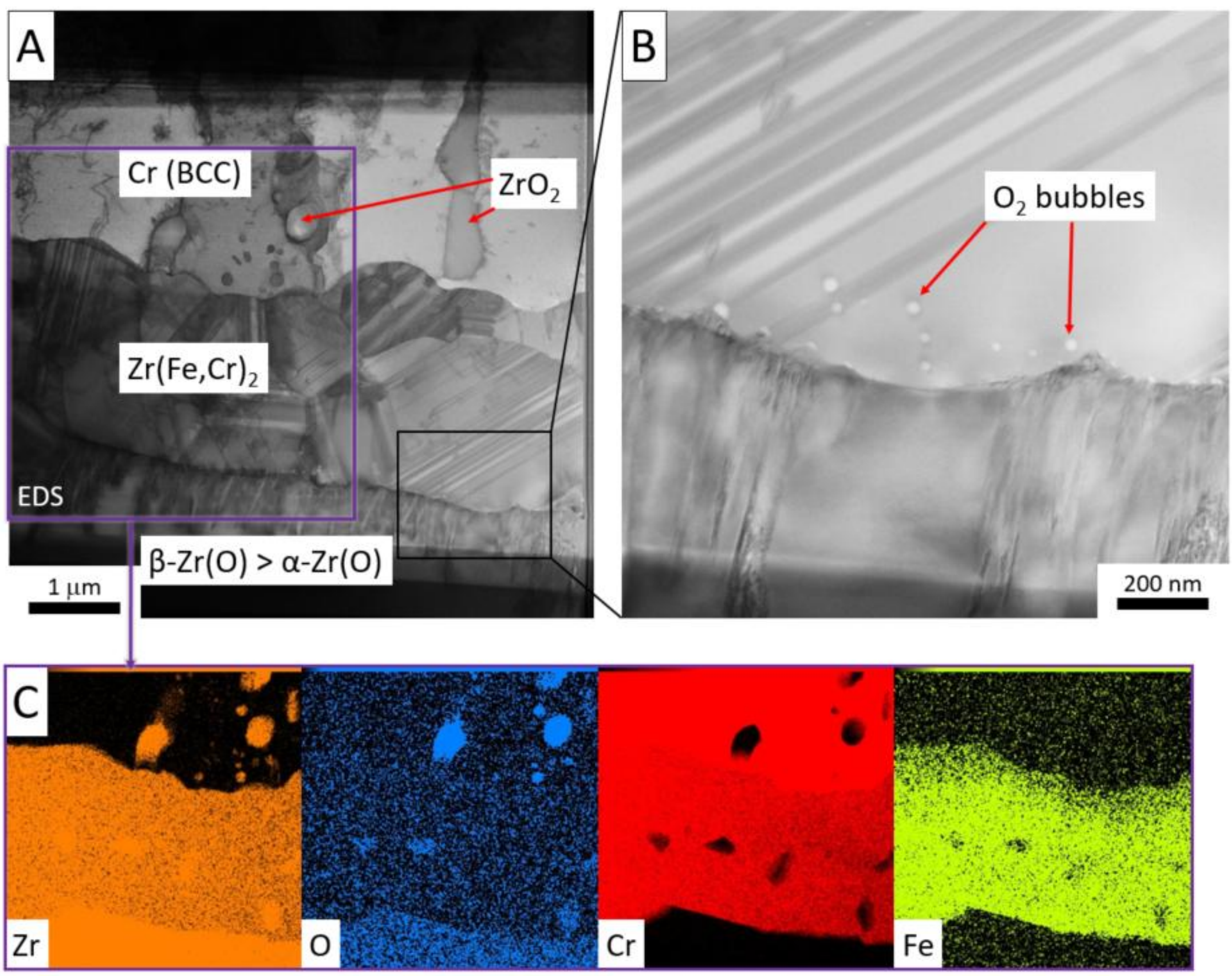

**Figure 8.** Close-up analyses of the Cr / $Zr(Fe,Cr)_2$ / Zr(O) interface: A) BF STEM image of the Cr / $Zr(Fe,Cr)_2$ / Zr(O) interfaces; B) Higher magnification BF STEM image of the nano-bubbles observed at $Zr(Fe,Cr)_2$ / Zr(O) interface; C) EDS maps (Zr, O, Cr, Fe) of the area marked with violet frame in (A), confirming higher Fe content in $Zr(Fe,Cr)_2$ layer.

**4. Conclusion**

In this work, a structural study of Cr-coated and uncoated Zircaloy-4 cladding material is presented. The structure of the surface layers was evaluated before and after high temperature X-ray diffraction (HT-XRD) studies, by detailed SEM and TEM analyses. Chromium coating deposited by AIP process acts as a barrier that is able to slow down oxidation of Zircaloy-4 (Z4) substrate. During HT-XRD experiment, formation of the $Cr_2O_3$ layer (T>600°C) is recorded at higher temperatures compared to formation of $ZrO_2$ (T>200°C), offering significant improvement over bare cladding material. At elevated temperatures, α-to-β transformation of Zr promotes Fe segregation, formation of $Zr(Fe,Cr)_2$ Laves phase and nano-bubbles at the former Cr / Z4 interface. Future studies may involve further evaluation of thermal/mechanical properties and irradiation studies of Cr-coated Zr-based cladding material, as well as exploration of different coatings compositions.

**Declaration of competing interest**

The authors declare that they have no known competing financial interests or personal relationships that could have appeared to influence the work reported in this paper.

**Acknowledgements**

This project has been supported by NCBiR grant ATF Cladding, grant agreement DWM/POLKOR/2/2024. This work was supported by the Korea Institute of Energy Technology Evaluation and Planning (KETEP) and the Ministry of Trade, Industry & Energy (MOTIE) of the Republic of Korea (RS-2023-00303582).

**Data Availability Statement**

Data will be made available on request.

**Supporting Information**

Supporting Information is available from the Wiley Online Library or from the author.